# GAS-SOLID COEXISTENCE IN HIGHLY CHARGED COLLOIDAL SUSPENSIONS

P.S. Mohanty[1], B.V.R. Tata[1], A. Toyotama[2] and T. Sawada[2]

[1]*Materials Science Division, Indira Gandhi Centre for Atomic Research, Kalpakkam -603 102, Tamil Nadu, India*
[2]*National Institute for Materials Science, 1-1 Namiki, Tsukuba, Ibaraki 305-0044, Japan*

Aqueous suspensions of highly charged polystyrene particles with different volume fractions have been investigated for structural ordering and phase behavior using static light scattering (SLS) and confocal laser scanning microscope (CLSM). Under deionized conditions, suspensions of high charge density colloidal particles remained disordered whereas suspensions of relatively low charge density showed crystallization by exhibiting iridescence for the visible light. Though for unaided eye crystallized suspensions appeared homogeneous, static light scattering measurements and CLSM observations have revealed their inhomogeneous nature in the form of coexistence of voids with dense ordered regions. CLSM investigations on disordered suspensions showed their inhomogeneous nature in the form coexistence of voids with dense disordered (amorphous) regions. Our studies on highly charged colloids confirm the occurrence of gas-solid transition and are in accordance with predictions of Monte Carlo simulations using a pair-potential having a long-range attractive term [Mohanty and Tata, *Journal of Colloid and Interface Science* **2003**, 264, 101]. Based on our experimental and simulation results we argue that the reported reentrant disordered state [Yamanaka *et al Phys*. *Rev*. *Lett*. **1998**, 80, 5806 and Toyotama *et al Langmuir*, **2003**, 19, 3236] in charged colloids observed at high charge densities is a gas-solid coexistence state.

# 1. Introduction

Structural ordering in monodisperse charge stabilized colloidal suspensions can be tailored with ease by varying the range and the strength of the interparticle interaction[1,2]. The most dominant interparticle interaction among the like-charged colloidal particles in a homogeneous charged stabilized suspension is screened Coulomb repulsion and is given by Derjaguin-Landau-Verwey-Overbeek (DLVO) theory[3]. However, there have been several observations *viz.*, reentrant transition observed as a function of salt concentration[4], vapor-liquid transition[5,6], observation of stable voids[7-9], which suggested the existence of a long-range attraction in the effective interparticle interaction $U(r)$ of like-charged colloidal particles contradicting the predictions of DLVO theory. The DLVO theory is advanced by Sogami-Ise by considering the counterion mediated attraction and regarding the macroionic part as a one-component system[10]. The effective pair-potential $U_s(r)$ obtained by Sogami and Ise[10] by considering Gibbs free energy, is found to have a long-range attractive term in addition to the usual screened Coulomb repulsive term given by DLVO theory. Tata *et al*'s computer simulations using $U_s(r)$ could explain the above-mentioned experimental observations satisfactorily[11-13] and alternative explanation, based on volume term theory which does not require an attractive term in the effective pair-potential also exists[14-15]. However, recent calculations based on Poisson Boltzmann (PB) cell model[16,17] have shown that the spinodal instability that arises in volume term theory is spurious and is due to linearization of PB equation. Further, our own pair-potential measurements on very dilute suspensions of highly charged colloidal particles have shown the existence of long-range attractive term in the $U(r)$ of like-charged colloids[18].

Suspensions of particles having low effective surface charge density $\sigma_e$ are known to exhibit a homogenous fluid to homogenous crystal transition upon lowering the salt concentration[19].

Whereas suspensions of highly charged colloidal particles, under deionized conditions, are found to be in an inhomogeneous state in the form of a rare phase (voids) coexisting with a dense phase having glass-like disorder[8]. The glass-like disorder of the dense phase was confirmed using CLSM and ultra-small-angle-x-ray scattering studies[8]. Though Monte Carlo (MC) simulations have predicted existence of gas-solid coexistence in the form of voids coexisting with ordered regions[12] for charge density beyond a critical value, there have been no systematic experimental studies verifying this prediction. We report here the experimental verification of this theoretical prediction in deionized aqueous of highly charged polystyrene colloids.

Further, Yamanaka et al[20] and Toyotama et al[21] have recently reported a reentrant solid-liquid transition (also known as reentrant order-disorder transition) by varying, $\sigma_e$ on deionized charged colloidal suspensions of silica and polystyrene particles respectively. Mohanty and Tata have performed detailed Monte Carlo (MC) simulation results[22] as a function of $\sigma_e$ and provided understanding for these experimental observations. These simulations have clearly shown that the reentered disordered state observed at high charge densities is a gas-solid coexistence state. Further, these simulations have revealed that the homogenous ordered phase exists over a narrow range of $\sigma_e$, whereas experiments[20,21] have indicated the existence of ordered region over much wider range of $\sigma_e$. We would like to point out that identification of ordered region in these experiments[20,21] is completely based on iridescence from the samples and no detailed investigations have been reported to characterize the structural ordering and for their inhomogeneous/homogenous nature. Motivated by these studies, we have carried out systematic static light scattering and CLSM studies on deionized aqueous suspensions of polystyrene particles for different surface charge densities. We report here the observation of gas-solid coexistence in the form of voids coexisting with ordered regions in deionized suspensions with particles of intermediate charge density and voids coexisting with glass-like

(amorphous) disordered dense regions at relatively high charge densities. Our results thus confirm the predictions of MC simulations that a homogenous crystalline state exists in a narrow range of charge density. This crystalline state becomes inhomogeneous in the form of gas-ordered (crystal) coexistence at intermediate values of $\sigma_e$ and gas-disordered (amorphous) coexistence at relatively high values of $\sigma_e$. We have also carried out MC simulations close to experimental parameters and our MC simulations results also showed gas-solid coexistence in agreement with experimental results.

## 2. Sample Details and Methods

**2.1. Sample Details.** Monodisperse polystyrene latex particles synthesized by Dr. Yoshida of Hashimoto Project, ERATO, JST[21] are used in the present investigations and the parameters of the suspensions are summarized in Table 1. These suspensions are purified by dialysis and ion-exchange method. After purification, the ion-exchange resin beads (AG501-X8, Bio-Rad Laboratories, Hercules, CA) were introduced into the suspensions and the suspensions were kept undisturbed for one week for deionization. The volume fraction of these suspensions was determined by drying up method[23]. The effective surface charge density $\sigma_e$ is determined by conductivity method[1,21].

Samples S1 to S7 are prepared by diluting the mother suspensions with appropriate amount of $H_2O$ and $D_2O$ mixtures such that the density of polystyrene particle matches with the density of the medium (50:50 in $H_2O$-$D_2O$ mixtures). The conductivity of $H_2O$ and $D_2O$ used for the dilution is less than 1 μS/cm. Samples for static light scattering studies were prepared in cylindrical quartz cells with 8mm optical path length. For further deionization, known amount of mixed-bed ion exchange resins were added to the samples. The cells were sealed hermetically and were left undisturbed for about a week.

**2.2. Static Light Scattering.** The structural ordering in deionized suspensions is characterized by static light scattering. The scattered intensity $I_s(Q)[= AP(Q)S(Q)]$ is recorded as a function of scattering wave vector $Q$. Structure factor $S(Q)$ as a function of $Q$ is calculated after correcting for particle structure factor $P(Q)$[24,25]. Where $Q = (4\pi\mu_m/\lambda)sin(\theta/2)$ is the scattering wave vector, $\theta$ is the scattering angle, $\mu_m$ is the refractive index of the dispersion medium and $\lambda$ is the wavelength of laser light. The particle scattering form factor $P(Q)$ for a spherical particle of radius '$a$' is given as,

$$P(Q) = \left\{ \frac{3[\sin(Qa)-(Qa)\cos(Qa)]}{(Qa)^3} \right\}^2 \qquad (1)$$

**2.3. Confocal Laser Microscopy.** The internal structure (homogeneous / inhomogeneous) of the deionized samples (S1-S7) was characterized using M/s Leica TCS SP2 RS, Germany inverted type confocal laser scanning microscope (CLSM). The sample cell used for CLSM studies is shown schematically in Figure 1. The cell is made up of cylindrical quartz tube of 8 mm in diameter and 25 mm in height. The cover glass is fixed to the polished end of the quartz tube by applying adhesive for outer surface of the tube. Deionized suspension with known volume fraction is introduced in the cell and a clean nylon bag containing the mixed bed of ion-exchange resins is hung from the top for further deionization. The top of the cell is sealed hermetically using Para-film. The cell is then mounted on CLSM stage for observations.

**2.4. Monte Carlo Simulations.** In order to understand the experimental observations, Monte Carlo (MC) simulations have been carried out using the Metropolis algorithm[26] with periodic boundary conditions for a canonical ensemble (constant $N$, $V$, $T$ where $N$, $V$, and $T$ are respectively, the number of particles, the volume, and the temperature) close to the experimental parameters. Particles are assumed to interact via a pair potential $U_s(r)$ having the functional form

$$U_s(r) = \frac{Z^2 e^2}{2\varepsilon} \left(\frac{\sinh(\kappa a)}{\kappa a}\right)^2 \left(\frac{A}{r} - \kappa\right) \exp(-\kappa r) \tag{2}$$

Where $A = 2[1 + \kappa a \coth(\kappa a)]$ and the inverse Debye screening length $\kappa$ is given as

$$\kappa^2 = 4\pi e^2 (n_p Z + C_s)/(\varepsilon k_B T), \tag{3}$$

Where $Ze$ is the effective charge on the particle (related to the surface charge density by $\sigma_e = Ze/\pi d^2$), $C_s$ the salt concentration, $T$ the temperature (298 K), $\varepsilon$ the dielectric constant of water, and $k_B$ the Boltzmann constant. The position of the potential minimum $R_m$ is given as $R_m = \{A + [A(A + 4)]1/2\}/2\kappa$ and its depth $U_m = U(R_m)$. Both $R_m$ and $U_m$ depend on $\sigma_e$ and $C_s$. Simulations have been carried out using the procedure reported in our earlier studies[22].

### 3. Results and Discussion

**3.1 Experimental Results.** One-week old samples S1 to S3 showed iridescence for the visible light over the entire volume of the suspension, whereas S5 to S7 did not exhibit any iridescence even after a month. Iridescence in samples S1 and S3 and absence of iridescence in sample S7 are shown in Figure 2, which are arranged in increasing order of $\sigma_e$. These observations indicate that samples S1 to S3 are crystalline and S5 to S7 are disordered. However the information regarding the homogeneous or inhomogeneous nature of these crystalline samples (S1 to S3) and disordered samples (S5 to S7) cannot be known through visual observations. So SLS and CLSM studies are carried out to investigate nature of these samples.

Figure 3 shows the measured $S(Q)$ for the samples S1 and S3. The peak positions in $S(Q)$ correspond to a body centered cubic (BCC) structure. Sample S2 has also shown BCC ordering. Assuming the samples are homogeneous, we have estimated the first peak position $Q_{max}$ from the volume fraction of the samples S1- S3 using the relation[8,23],

$$\phi = \frac{\pi d^3}{6} \frac{1}{\sqrt{2}} \left(\frac{Q_{max}}{2\pi}\right)^3 \qquad (4)$$

The position of $Q_{max}$ is shown as dotted line in Figure 3. Notice that the first peak in $S(Q)$ occurs at higher value of $Q$ than at $Q_{max}$ in both the samples S1 and S3. Similar shift also has been observed for sample S2. The volume fraction ($\phi_d$) estimated from the peak position in $S(Q)$ is found to be larger than $\phi$ and the corresponding interparticle separation is found to be smaller than the average interparticle separation $d_o$ ($d_o = (\sqrt{3}/2)[\pi d^3/3\phi]^{1/3}$). These observation imply that the crystallized samples, which appeared homogeneous for unaided eye, are inhomogeneous (i.e. the ordered phase does not occupy the full volume of the medium). The fraction of the volume that is not occupied by the ordered region is expected to appear as particle free regions (voids). The estimated void fraction $V_f$ (= [1- ($\phi/\phi_d$)]) for samples S1, S2 and S3 are 0.27, 0.47 and 0.36 respectively. Multiple scattering in samples S5-S7 is found to be considerable which prevented us to measure $S(Q)$ in these samples.

CLSM studies have been carried out in the same samples by transferring the suspensions (S1-S3) from light scattering cells to CLSM cells. Figure 3 shows the frame averaged (averaged over 20 frames) CLSM images in samples S1 and S3 at a depth of 55 μm and 70 μm respectively from the cover slip. These images clearly show that the crystallites (dense phase) coexist with voids (rare phase). Sample S2 also showed coexistence of crystallites with voids. Thus CLSM observations confirm unambiguously the inhomogeneous nature of crystalline samples as revealed by static light scattering.

Further, CLSM measurements samples S2 and S3 have revealed that void fraction in increases with decreasing $\phi$. This implies that in samples which exhibited inhomogeneous nature, the ordered phase is majority phase and the voids (rare phase) is a minority phase. The volume fraction of minority phase (*i.e.* void fraction) increases as $\phi$ decreases. Below certain $\phi$ the voids form the majority phase and the suspension is expected to exhibit macroscopic phase separation ( *i.e.* phase separation observable by the unaided eye). Indeed we observed in a dilute sample S4, macroscopic phase separation (Inset photograph in Figure 4) in the form of dense phase at the bottom of the cell having iridescence and a rare phase at the top. *S(Q)* measured in the top region revealed that the rare phase is a gas-like disorder (Figure 4). The appearance of macroscopic phase separation at lower volume fraction can be understood as follows. At higher $\phi$, the ordered phase is the majority phase and is connected. So the suspension looks homogeneous to the unaided eye at higher $\phi$ (= 0.001 to 0.005) and the rare phase (gas-like ordered) is the minority phase which appear as voids within the dense phase. Below a certain volume fraction (~ 0.001), the voids constitute majority phase because void fraction is high. The ordered phase is the minority phase and appears as clusters [9] during phase separation. Since the density of these dense phase cluster is more than the density of the solvent, these clusters settle down due to gravity and coalesce leading to a macroscopic phase separation as shown in Figure 4.

CLSM studies also have been carried out on samples S6, S7 which did not exhibit iridescence but appeared homogeneous for the unaided eye. Figure 5 shows the CLSM image of the sample S7 taken at a depth of 60 μm from the cover slip. From the image, it can be seen that the voids coexist with dense phase which is disordered. The disorder with in the dense phase is characterized to be glass-like (amorphous) by performing averaging over frames. If the disorder is solid-like the image obtained by frame averaging is expected to be much sharper than the single image. We found that CLSM image obtained by averaging over 20 frames has improved the sharpness of the image, which suggests that the dense disordered regions are glass-like (amorphous)[8,9] Sample S6 also showed similar behavior. These observations on samples of high charge density particles corroborate the earlier studies by Tata *et al*[8,9] on suspensions of highly charged poly(chlorostyrene-styrene sulfonate) particles. The dilute sample S5 of highly charged particles showed macroscopic phase separation in the form of dense disordered phase at the bottom of the sample cell and a rare phase at the top of the cell, similar to that observed in sample S4.

Thus our studies using static light scattering and CLSM on deionized suspensions of highly charged particles confirm the occurrence gas-solid coexistence in the form of voids coexisting with ordered regions at intermediate charge density and voids with disordered regions at higher charge density.

**3.2. MC Simulation Results.** Occurrence of gas-solid transition in highly charged colloids suggest the existence of long-range attraction in the effective pair-potential. Hence $U_s(r)$ is chosen as the effective pair-potential and MC simulations have been carried out close to the experimental parameters to understand the experimental observations. Here we present the simulations results for suspension parameters close to samples S3, S7 and S4. Figure 6A shows that the pair-correlation function *g(r)* for the suspension S3. The peaks in *g(r)* correspond to bcc crystalline order. The dotted vertical line shows the position of the average interparticle separation, $d_o$ calculated for a homogeneous suspension at $\phi = 0.005$. Notice that the first peak of *g(r)* occurs at distance less than '$r/d_o$', implying that the particles in the suspension do not occupy the full volume of the MC cell. The corresponding projected particles in the MC cell (Figure 6B) show particle-free regions (voids) coexisting with ordered regions. Thus MC simulation results for sample S3 agree with experimental results of SLS and CLSM studies. Simulations on samples S1 and S2 also showed gas-crystal coexistence which is in agreement with experiments. .

The pair-correlation function $g(r)$ correspond ding to sample S7 shows (Figure 6C ) shows a split second peak and decays as a function of $r$. This suggests glass-like disorder in the suspension. Further the first peak position occurs at distance smaller than $r/d_o$ indicating that the disordered region does not occupy the full volume of the MC cell. The corresponding projected particles in the MC simulation cell (Figure 6D) showed particle free regions (voids) coexisting with dense disordered regions. The disorder in the dense phase is identified to be glass-like from the spilt second peak in $g(r)$ and from the saturation behavior of mean square displacement (MSD) (Inset of Figure 6C). Simulations for sample S6 also coexistence of voids with dense disordered (glass-like) regions. This MC simulation results are in agreement with the CLSM observations on sample S6 and sample S7.

Figure 6E shows $g(r)$ as a function of $r/d_o$ for sample S4. The peaks in $g(r)$ and the gradual decay of $g(r)$ to 1 as function of $r/d_o$ suggest existence of dense phase clusters with a liquid-like structural ordering within the cluster[11]. Further, the first peak position in $g(r)$ is close to the $R_m$ value of the $U_s(r)$. The corresponding projections of particles in MC cell (Figure 6F) confirm the existence of dense disordered clusters . In an experimental situation, these clusters settle down due to gravity, leading to a macroscopic phase separation in the form of dense phase coexisting with a rare phase. The above simulation results can be understood from the behavior of $U_s(r)$ on $\sigma_e$ and $\phi$.

Figure 7A shows the potential $U_s(r)$ for values of $\sigma_e$ = 0.4, 0.51 µC/cm² at $\phi$ = 0.005. The vertical line shows $d_o$ corresponding to $\phi$ = 0.005. If $R_m$ is less than $d_o$ and for well depth $U_m \gg k_BT$ (Table 1), particles which are initially at a distance of $d_o$ experience strong attractive forces and get trapped into such deep wells with a new interparticle spacing close to $R_m$ leading to an inhomogeneous suspension. As the total volume of the suspension remains unchanged, this also leads to the appearance of voids. If the difference between $R_m$ and $d_o$ is small ($\approx 0.04 d_o$, see Figure 7A, curve ' a ', Table 1), the dense phase appears as crystalline. Note that if the difference between $R_m$ and $d_o$ is more ($\approx 0.2 d_o$, Figure 7A, curve ' b ', Table 1) and well depths are large, then the particles suffer large random displacements due to the strong attractive forces and get arranged randomly during the dense phase formation, resulting in an amorphous structure in these suspensions. When this difference between $R_m$ and $d_o$ is sufficiently large ($> 0.3 d_o$, see Figure 7B), the dense phase is expected to be initially in the form of clusters as seen in Figure 6F. In an experimental situation, such clusters sediment due to gravity leading to a macroscopic phase separation.

**3.3 Discussion.** As mentioned earlier Yamanaka *et al*[20] and Toyotama *et al*[21] have reported that initially a homogenous disordered (liquid-like ordered) deionized suspension undergoes crystallization upon increasing $\sigma_e$ on the particles. This crystalline order (solid phase) is found to disorder once again on further increase of $\sigma_e$. This reentrant order-disorder transition (also referred to as reentrant solid-liquid transition) is understood using MC simulations[24] with $U_s(r)$ as the pair-potential. These simulations showed existence of homogenous ordered phase in a narrow range of $\sigma_e$ whereas the solid region reported by Yamanaka *et al*[20] and Toyotama *et al*[21], which is based on iridescence, existed relatively over a wide range of $\sigma_e$. Present investigations using static light scattering and CLSM have reveled that for $\sigma_e$ values in the range of 0.24 – 0.4 µC/cm$^2$ the suspensions do not exhibit homogenous crystalline state but are in an inhomogeneous state and it corresponds to gas-crystal coexistence. This gas-crystal coexisting state becomes inhomogeneous and disordered upon increasing $\sigma_e$ beyond 0.4 µC/cm$^2$. Yamanaka *et al*[20] and Toyotama *et al*[21] have identified this inhomogeneous disordered region as reentrant liquid region. However, present studies have clearly revealed that the reentered liquid region is also a gas-solid coexistence state where the structural ordering of particles in the solid corresponds to a glass-like disorder. Thus we conclude that phase diagrams reported by Yamanaka *et al*[20] and Toyotama *et al*[21] consists of a homogenous liquid-like region for low values of $\sigma_e$ (< 0.2 µC/cm$^2$) and a homogenous crystalline region for $\sigma_e$ < 0.24 µC/cm$^2$. An inhomogeneous crystalline region exists in the form of voids coexisting with dense ordered regions up to $\sigma_e$ ≤ 0.4 µC/cm$^2$. For charge densities beyond this $\sigma_e$ deionized suspensions will be in an inhomogeneous disordered state which will be in the form of coexistence of voids with a dense phase having glass-like disorder. Hence we conclude that a dilute charged colloidal suspension under deionized conditions exhibit a homogeneous liquid to homogenous crystal transition upon increasing $\sigma_e$. This homogenous crystalline state once again becomes inhomogeneous in the form of a gas-solid coexistence state upon further increase of $\sigma_e$. Present results also suggest that the reentrant liquid or reentrant disordered state reported by Yamanaka *et al*[20] and Toyotama *et al*[21] is a gas-solid coexistence state and suspensions of highly charged colloidal particles exhibit gas-solid transition upon deionization (i.e. lowering the salt concentration).

## 4. Conclusions

For the first time, an inhomogeneous crystalline phase in the form of voids coexisting with ordered regions in deionized dilute charged colloidal suspensions have been investigated using static light scattering and confocal laser scanning microscope. This inhomogeneous crystalline region occurs in relatively narrow range of $\sigma_e$. For charge densities beyond this range suspensions remain in an inhomogeneous disordered state in the form of voids coexisting with glass-like disordered regions. Thus highly charged colloids exhibit gas-solid coexistence under deionized conditions. Occurrence of gas-solid transition in charged colloids provide a strong evidence for counter ion mediated long-range attraction in the effective pair-potential. Present results are expected to motivate theoreticians to propose microscopic mechanism for counterion mediation leading to long-range attraction between like-charged colloids.

**Acknowledgements:** We thank Dr. Junpei Yamanaka for fruitful collaboration and for useful discussions.

**TABLE CAPTION**

**Table 1.** Sample details (effective surface charge density $\sigma_e$, volume fraction $\phi$, diameter of the particle $d$, average interparticle separation $d_o$), parameters of pair-potential (position of the potential minimum $R_m$ and its depth $U_m$) and nature of the coexistence state as identified by static light scattering and confocal laser scanning microscope studies. Abbreviations G, C, MP and A represent gas, crystalline, amorphous and macroscopic phase separation respectively.

**FIGURE CAPTIONS**

**Figure 1:** Schematic diagram of sample cell used for CLSM investigations. (1) Cover glass, (2) Adhesive, (3) Suspension and (4) Nylon bag containing mixed bed of ion exchange resin.

**Figure 2:** Photographs of the samples S1, S3 and S7 arranged in increasing order of $\sigma_e$. Samples S1 and S3 show iridescence, whereas sample S7 shows no iridescence. Mixed bed of ion-exchange resin can be seen at the bottom of the sample cells.

**Figure 3:** Structure factor $S(Q)$ vs scattering wave vector $Q$ for samples S1 and S3. The vertical dotted lines shown in the figures correspond to $Q_{max}$ calculated from the Eq. 4 for homogeneous suspensions. The arrows (from left to right) in figures S1 and S3 represent the peak positions corresponding to Bragg reflections [(110),(200),(211),(220),(222),(310),(321),(330),(400),(411),(420),(422)] and [(110),(200),(211)] of the dense crystalline phase respectively. These reflections characterize the crystalline state as body centered cubic (BCC). The corresponding CLSM images for samples S1 and S3 taken respectively at distance of 55 μm and 70 μm from the cover glass show coexistence of voids with dense phase (crystallites). Images are taken using wavelength $\lambda$ = 488 nm of Ar-ion laser and a 40× /0.75 objective. Scale bar = 20 μm.

**Figure 4:** Photograph of the sample S4 (shown as an inset) exhibiting macroscopic phase separation in the form of dense phase at the bottom of the cell coexists with a rare phase at the top. Mixed bed ion-exchange resins can be seen at the bottom of the cell. The iridescence for the visible light from the dense phase implies that it has crystalline order. $S(Q)$ vs. $Q$ measured in the rare phase shows that the rare phase has gas-like disorder.

**Figure 5:** CLSM image after averaging over 20 frames shows coexistence of voids with dense disordered (amorphous) regions for the sample S7 taken at a distance of 60 µm from the cover glass. Images are taken using $\lambda$= 488 nm of Ar-ion laser and a 40× /0.75 objective. Scale bar = 20 µm.

**Figure 6:** Pair-correlation function $g(r)$ vs $r/d_o$ and the corresponding projection of the particle coordinates for sample S3, S7 and S4 are shown in A, B; C, D; and E, F respectively. The vertical dotted lines in A, C and E correspond to $d_o$ for homogeneous suspensions with $\phi$ = 0.005, 0.005 and 0.0005, respectively.

**Figure 7:** Pair-potential $U_s(r)/k_BT$ vs $r/d_o$ for (A) sample S3 (curve 'a'), sample S7 (curve 'b') and (B) for sample S4. The vertical dotted lines shown in (A) and (B) correspond to $d_o$ for homogeneous suspensions with $\phi$ = 0.005 and $\phi$ = 0.0005 respectively.

**Table 1**

| Sample No. | $\sigma_e$ ($\mu C/cm^2$) | $\phi$ | $d$ (nm) | $d_o/d$ | $R_m/d$ | $U_m/k_BT$ | Coexistence State |
|---|---|---|---|---|---|---|---|
| S1 | 0.24 | 0.001 | 136 | 8.8 | 6.86 | -1.85 | G + C |
| S2 | 0.4 | 0.001 | 104 | 8.8 | 7.8 | -2.04 | G + C |
| S3 | 0.4 | 0.005 | 104 | 5.143 | 5.146 | -2.97 | G + C |
| S4 | 0.4 | 0.0005 | 104 | 11.08 | 8.83 | -1.82 | MP (C + G) |
| S5 | 0.51 | 0.0005 | 126 | 11.08 | 7.28 | - 6.3 | MP (A+G) |
| S6 | 0.51 | 0.001 | 126 | 8.8 | 6.56 | -6.94 | G + A |
| S7 | 0.51 | 0.005 | 126 | 5.143 | 4.23 | -10.04 | G + A |

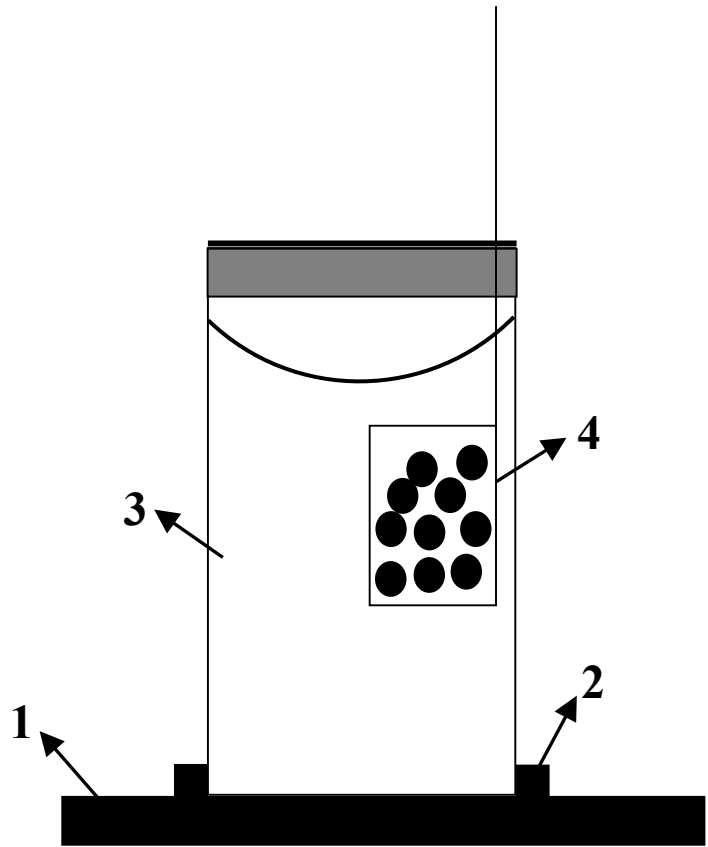

Figure 1: Mohanty et al

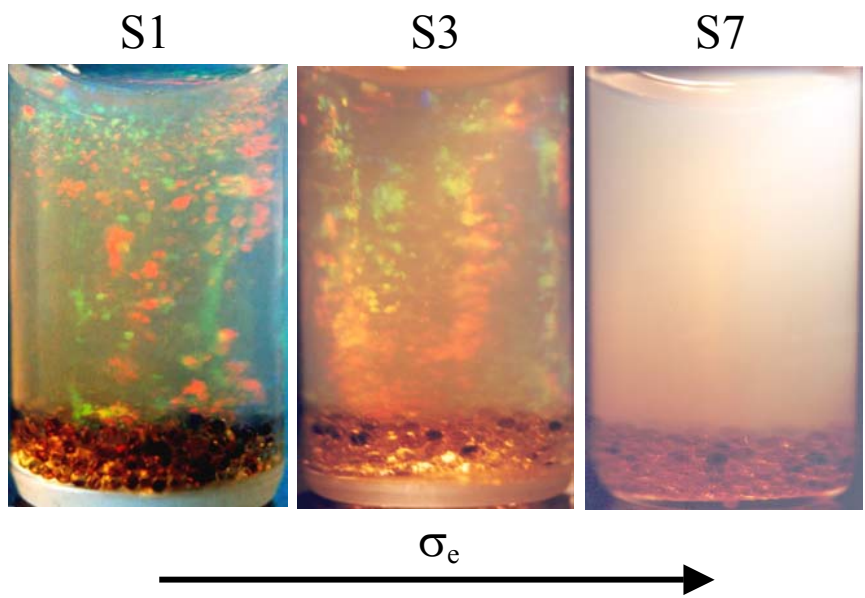

Figure 2: Mohanty et al

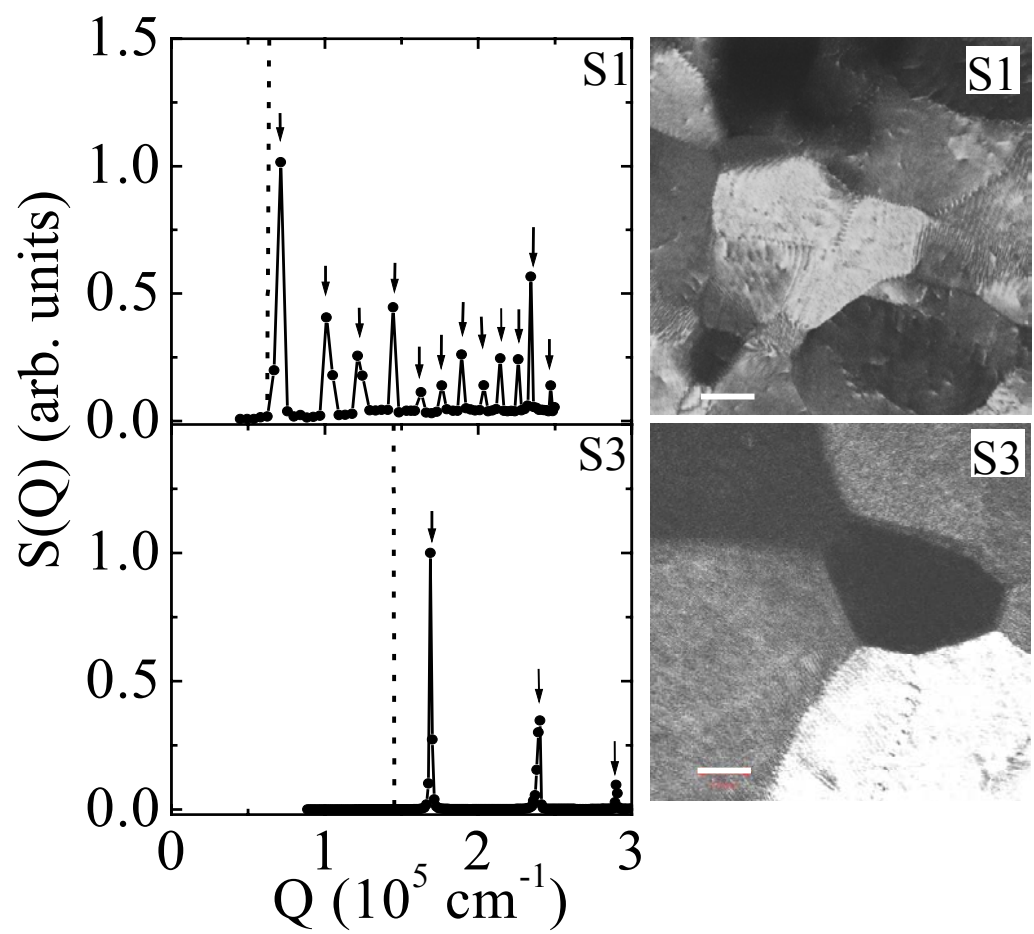

Figure 3: Mohanty et al

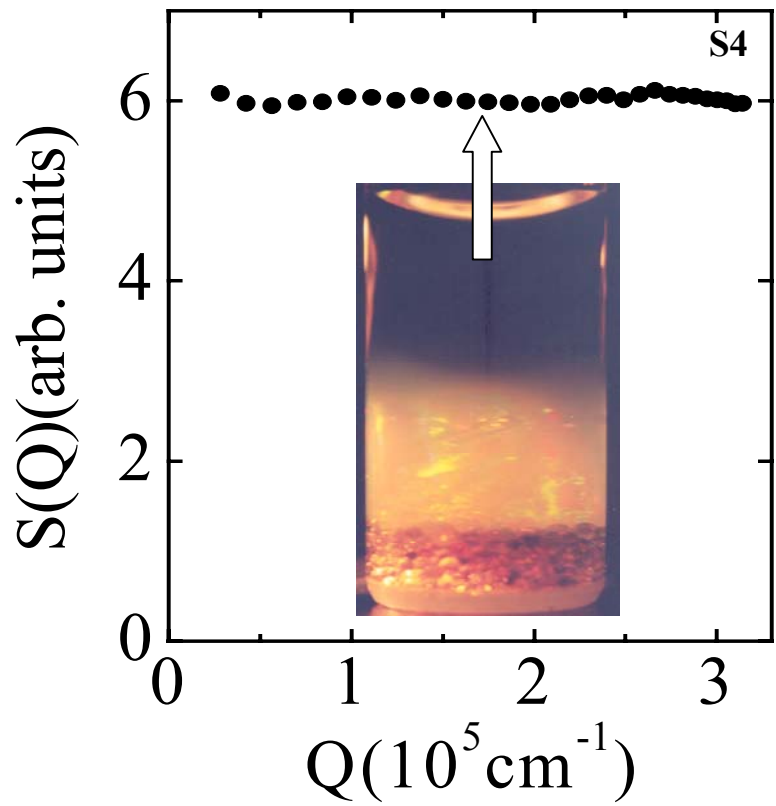

Figure 4: Mohanty et al

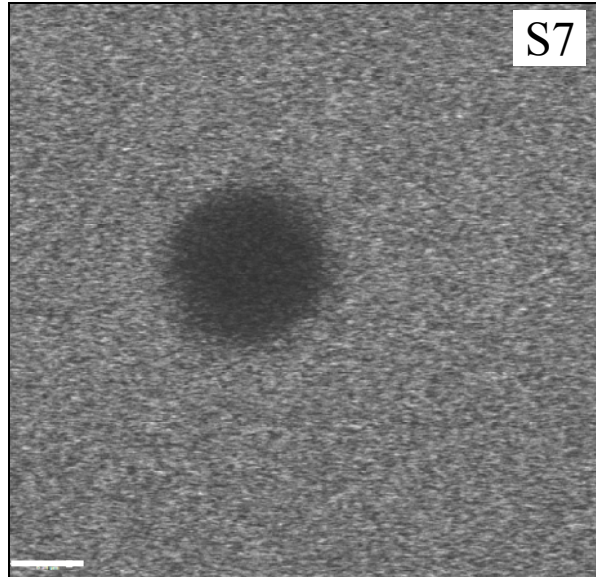

Figure 5: Mohanty et al

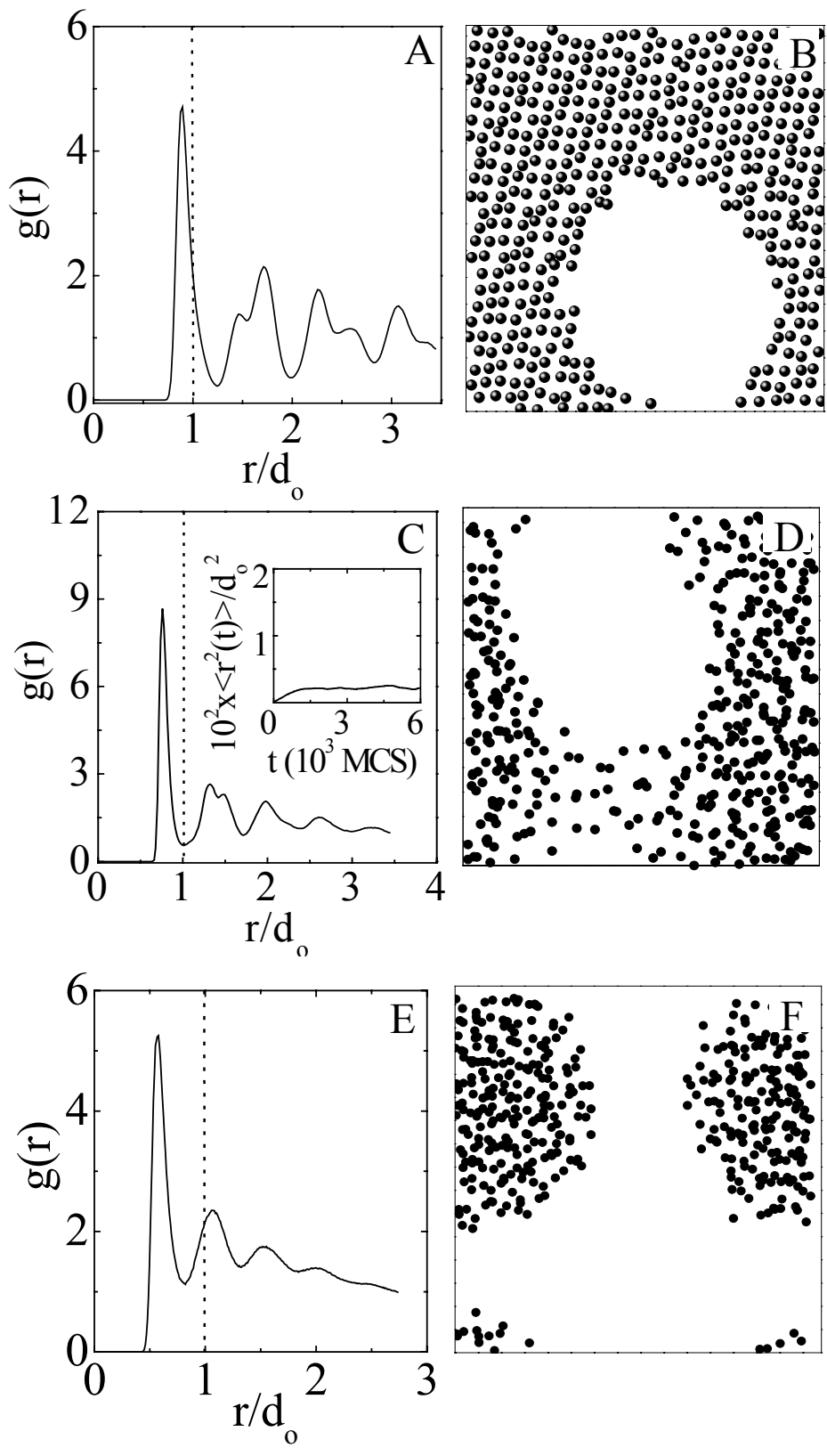

Figure 6: Mohanty et al

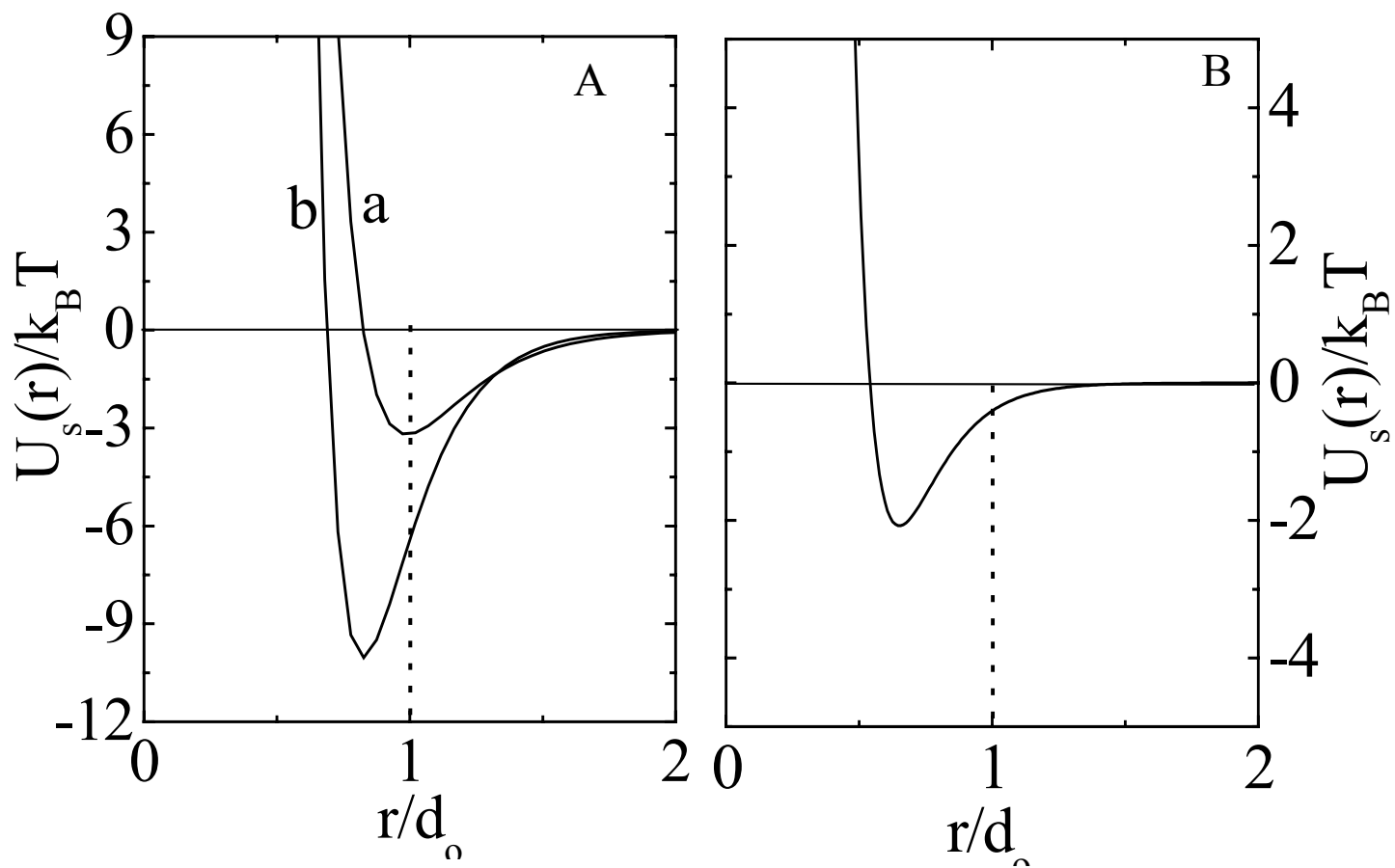

Figure 7: Mohanty et al